%% file: ISMIR2022_template.tex
\documentclass{article}
\usepackage[T1]{fontenc} 
\usepackage[utf8]{inputenc} 
\usepackage{ismir,amsmath,cite,url}
\usepackage{graphicx}
\usepackage{color}
\usepackage{multirow}
\usepackage{float}

\usepackage{fancyvrb}
\usepackage{fvextra}

\usepackage{lineno}

\title{A Dataset for Greek Traditional and Folk Music: L\lowercase{yra}}

\multauthor
{Charilaos Papaioannou$^1$, Ioannis Valiantzas$^2$, Theodoros Giannakopoulos$^3$,} { \bfseries{Maximos Kaliakatsos-Papakostas$^4$, Alexandros Potamianos$^1$}\\
$^1$ School of ECE, National Technical University of Athens, Greece\\
$^2$ Department of Music Studies, National and Kapodistrian University Of Athens, Greece\\
$^3$ National Center for Scientific Research - Demokritos, Greece\\
$^4$ Athena RC, Greece\\
{\tt\small cpapaioan@mail.ntua.gr}
}

\def\authorname{C. Papaioannou, I. Valiantzas, T. Giannakopoulos, M. Kaliakatsos-Papakostas and A. Potamianos}

\begin{document}

\maketitle
\begin{abstract}

Studying under-represented music traditions under the MIR scope is crucial, not only for developing novel analysis tools, but also for unveiling musical functions that might prove useful in studying world musics. This paper presents a dataset for Greek Traditional and Folk music that includes 1570 pieces, summing in around 80 hours of data. The dataset incorporates YouTube timestamped links for retrieving audio and video, along with rich metadata information with regards to instrumentation, geography and genre, among others. The content has been collected from a Greek documentary series that is available online, where academics present music traditions of Greece with live music and dance performance during the show, along with discussions about social, cultural and musicological aspects of the presented music. Therefore, this procedure has resulted in a significant wealth of descriptions regarding a variety of aspects, such as musical genre, places of origin and musical instruments. In addition, the audio recordings were performed under strict production-level specifications, in terms of recording equipment, leading to very clean and homogeneous audio content. In this work, apart from presenting the dataset in detail, we propose a baseline deep-learning classification approach to recognize the involved musicological attributes. The dataset, the baseline classification methods and the models are provided in public repositories. Future directions for further refining the dataset are also discussed.
 
\end{abstract}

\input{1_intro}

\input{2_dataset}

\input{3_application}

\input{4_results}

\input{5_discussion}

\input{6_conclusions}

\input{7_acknowledgements}

\bibliography{ISMIRtemplate}

%
%
%
%
%

\end{document}

%% file: 1_intro.tex
\section{Introduction}\label{sec:intro}


Traditional music of Greece is under-represented in available datasets, despite the fact that this music offers unique perspectives that combine characteristics of Western and Eastern music. The development of a traditional Greek music dataset is interesting to study in its own right, while it is also expected to provide a more complete picture of the music in the Mediterranean and Europe. Computational methods have been extensively studied in the past decades for carrying out ethnomusicological studies, constituting the field of Computational Ethnomusicology. A review of such methods is presented in~\cite{panteli2018review}, where it is evident that there are many benefits in compiling datasets of traditional music that can readily be used in computational models.

Several such datasets have been developed. For instance, an enormous database of Dutch melodies and songs that allows studying multiple musicological aspects is presented in~\cite{van2019documenting}. Similarly, a dataset of Indian art music was presented in~\cite{srinivasamurthy2014corpora}, where various MIR-related tasks were adjusted and applied therein. The employment of standard MIR tools has been evaluated in Arab-Andalusian music~\cite{repetto2018open} and Flamenco music with the COFLA dataset~\cite{kroher2016corpus}. Iranian Dastgah classification has been studied with MIR tools in~\cite{abdoli2011iranian}. A dataset of Georgian vocal music from historic tape recordings of three-voice songs was presented in~\cite{rosenzweig2020erkomaishvili}, where the aim was to preserve cultural heritage and also to apply and examine existing MIR methods (e.g., for pitch and onset detection) in data of this form. Such datasets also provide opportunities for studying idiosyncratic musical instruments, e.g., in the case of Chinese music~\cite{gong2021chmusic}, where many instruments are employed that are not related to the ones used in Western music.

In several cases, off-the-shelf MIR tools are not well-suited for tasks that include traditional musical types. For instance, studying melodic similarity of Turkish non equally-tempered makam phrases in symbolic format required the development of a novel representation based on MIDI~\cite{karaosmanouglu2014symbolic}. Similarly, rhythmic attributes of makam music with new models that integrate note transition rules for this music had to be developed for lyrics-to-audio alignment~\cite{dzhambazov2016use}. In addition, a novel method, based on audio signal processing, was created for identifying asymmetric rhythms in Greek traditional music~\cite{fouloulis2013traditional}.

The ``Lyra'' dataset of traditional and folk Greek music aims to contribute to all the aforementioned fields, from data-driven (computational) ethnomusicology to shaping new directions of research for MIR tools. The majority of available datasets that concern Greek music are unsuitable for computational analysis due to their unstructured nature. An exception is the Greek Audio Dataset (GAD)~\cite{makris2014greek}, which includes 1000 pieces with audio content (YouTube links), lyrics and annotations for genre (coarse categories including traditional, pop, rock), mood (valence-arousal plane coordinates) and pre-computed audio features; this dataset was later expanded in the Greek Music Dataset (GMD)~\cite{makris2015greek}, consisting of 1400 audio pieces with the aforementioned data. However both the GMD and the GAD are not focused on traditional and folk music, while the quality of audio recordings is varying significantly between genres.

The Lyra dataset presented in this paper, in contrast to the GMD, includes mainly traditional and folk Greek music with fine-grained labeling, focusing on musicological aspects of interest. Additionally, the recording quality is homogeneous across all pieces. Musicological soundness and high quality content are ensured by the fact that data has been collected and annotated from a documentary series that was presented by academics in Greek television. Some baseline classification tasks that are of particular interest in this dataset are presented, namely classification of pieces in genres, instruments and places of origin. In all cases, the results show that computational analysis can provide useful insight about musicological relations and phenomena in traditional and folk music of Greece -- and, possibly, of other places.


%% file: 2_dataset.tex
\section{Dataset Extraction and Description}\label{sec:dataset}


\subsection{Challenges and Methods}\label{subsec:method}



Large amounts of clean data is fundamental for current AI models to achieve their full potential. In this paper, we walked all the way, from the ``data in-the-wild'' multimedia content of a TV show to a fully annotated dataset, through a combination of machine automation and human evaluation/annotation processes. The consistency of the dataset and the richness of information it provides are tested by developing and training models that perform three different classification tasks. 

In the case of Greek traditional and folk music, there are few cases where metadata is combined with recordings in a structured manner. Additionally, there is a matter of quality of recordings as it is significantly affected by various factors, including the equipment used, the social occasion (e.g., during a festival or inside a studio) and the time period in which it took place, i.e., older recordings tend to be of lower quality.


An integration of dissimilar recordings, in terms of quality, can introduce significant deficiencies towards studying the musicological characteristics of world music with computational tools. In order to truncate the effect of the audio quality factor, we decided to incorporate the episodes from the Greek documentary series "To Alati tis Gis - Salt of the Earth" broadcasted by ERT (Hellenic Broadcasting Corporation), where primarily traditional and folk music is presented. The episodes were filmed during a 10 year period under strict production-level specifications, resulting to very clean and homogeneous audio content while significant wealth of information is provided by the presenter and the guests in the form of narrations between music performances.

The presented dataset consists of both the multimedia content and the annotations of interest. The multimedia content is provided as start and end timestamps that correspond to a single music piece, as parts of a longer episode, which is available online. Regarding the annotations, a taxonomy of labels is defined, based on the potential purposes of studies that might involve this dataset, considering also what metadata information can be retrieved either directly from the source or be integrated by volunteer annotators during the data collection process.

The study of Greek traditional and folk music involves knowledge about (i) the instrumentation, (ii) the genres, (iii) the places of origin and (iv) the way listeners perceive this music in terms of ``danceability'', among others. While musical instruments, genres and geography are semantically well-defined, the same can not be claimed for listeners' perception. Having at hand the multimedia content, i.e., audio and video, can be helpful to this end. Annotation about whether a music piece is being danced during its live performance can reveal cultural characteristics regarding the way this piece is perceived by the community, because body movements play an important role in music perception \cite{leman2015role}.

As a result, the taxonomy consists of (i) the musical instruments participating in the performance of each music piece (singing voice is considered an instrument), (ii) the musical genres and sub-genres that are identified by musicologists in Greek music, (iii) the places of origin and (iv) whether the music piece is being danced during its performance.


Volunteer annotators, students of the Department of Music Studies, undertook the task of separating each episode in music pieces and also labeling each one of them according to the specified taxonomy. A helper website was utilized where the respective category labels were added. An account was created for each annotator for the label assignment task. Every piece was labeled by two annotators and the final labels are the set of them where both annotators agree. At the end, the dataset that contains the aforementioned annotations along with the timestamps and the respective video id for each music piece was extracted from the database of the helper site.

\subsection{Dataset description}\label{subsec:data_description}

Lyra dataset is organized into a single table where each row corresponds to a music piece while the columns include the various metadata information. \tabref{tab:metadata} demonstrates the metadata categories.

\begin{table}
 \begin{center}
 \begin{tabular}{|c|c|c|c|}
  \hline
  \textbf{name} & \textbf{\# unique} &  \begin{tabular}{@{}c@{}}\textbf{multi-} \\ \textbf{label} \end{tabular} & \textbf{description} \\
  \hline
  id & 1570 & No & \begin{tabular}{@{}c@{}}unique identifier \\ of each \\ music piece \end{tabular} \\
  \hline
  instruments & 32 & Yes & \begin{tabular}{@{}c@{}}instruments \\ participating in \\ performance \end{tabular} \\
  \hline
  genres & 32 & Yes & \begin{tabular}{@{}c@{}}music style \\ annotations \end{tabular} \\
  \hline
  place & 81 & \begin{tabular}{@{}c@{}}Yes (to \\ contain \\ regions) \end{tabular} & \begin{tabular}{@{}c@{}}full hierarchy \\ of the place \\ of origin \end{tabular} \\
  \hline
  coordinates & 81 & No & \begin{tabular}{@{}c@{}}latitude and \\ longitude \end{tabular} \\
  \hline
  is-danced & 2 & No & \begin{tabular}{@{}c@{}}binary \\ value (0 or 1) \end{tabular} \\
  \hline
  youtube-id & 74 & No & \begin{tabular}{@{}c@{}} id of the \\ episode \\ available online \end{tabular} \\
  \hline
  start-ts & 1570 & No & \begin{tabular}{@{}c@{}} start timestamp \\ of the piece \end{tabular} \\
  \hline
  end-ts & 1570 & No & \begin{tabular}{@{}c@{}} end timestamp \\ of the piece \end{tabular} \\
  \hline
 \end{tabular}
\end{center}
 \caption{Metadata contained in the dataset.}
 \label{tab:metadata}
\end{table}

Beginning with the simplest metadata categories, in terms of description, ``id'' is a unique identifier for each piece, generated by its title, replacing Greek with Latin characters and spaces with dashes. As expected, the number of unique values will be the same with the number of pieces, namely 1570. The same stands for ``start-ts'' and ``end-ts'' that denote the exact time (second) that a song starts and ends in the corresponding video. The duration of the music pieces sums to approximately 80 hours.

The column ``youtube-id'' contains the id under which the video of the full episode is available online. The count of unique values are essentially the number of episodes that were used for the creation of the dataset. A typical duration of an episode is roughly a hundred minutes. The ``is-danced'' binary label informs about whether a music piece is being danced by the guests of the show. The music pieces annotated with ``1'' are approximately 51\% while in the rest of them no dance performance occurs.

\begin{figure}
 \centerline{\framebox{
 \includegraphics[width=0.8\columnwidth]{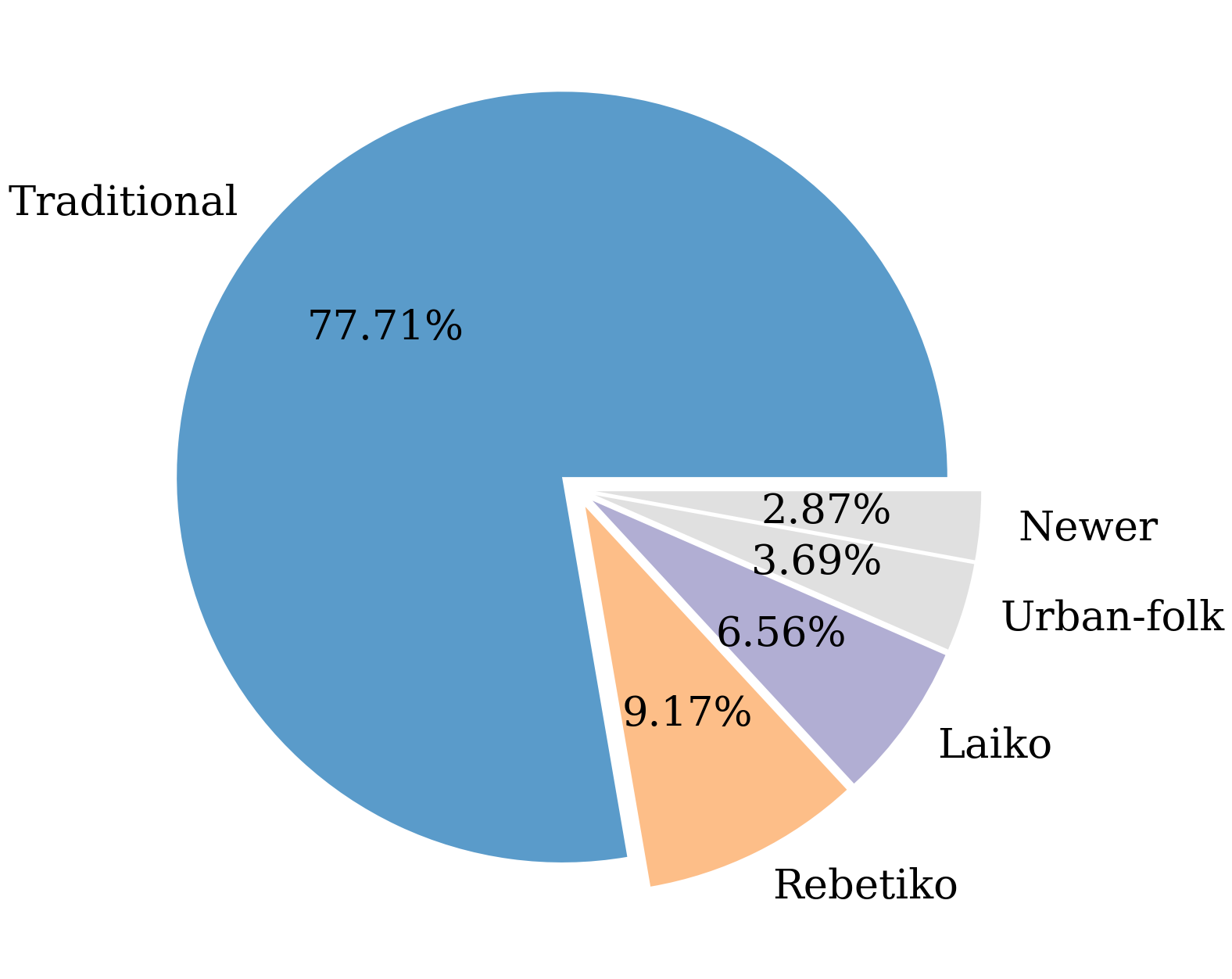}}}
 \caption{Relative frequencies of the music genres in the dataset.}
 \label{fig:genres}
\end{figure}

The classification of Greek music in ``genres'' is a work that requires one to take into account a number of socio-cultural and anthropo-geographical criteria. At an abstract level, we can distinguish the music of urban centers in contrast to the music of rural areas of Greece, with the former including rebetiko, laiko, urban-folk among others, while the latter, the music of rural areas, is what we generally call traditional music. 
\figref{fig:genres} shows the frequencies of the genres in the dataset, with ``traditional'' being the dominant one constituting almost 78\% of the total. Depending on the place of origin, the style of a traditional music piece varies accordingly and, thus, several sub-genres flourish, such as Epirotic for the songs originated from Epirus. The 32 unique values in this metadata category are separated into 5 distinct genres and 27 sub-genres.


From a musicological perspective, Greek traditional music can be divided into two large geographic areas, i.e., the island and the mainland Greece. Each one creates a distinctive musical feeling as there are large variations both on the rhythmic approach and the scales that are commonly used. For example, in islands we frequently come across music pieces with simple, fast rhythms while in the mainland more complex, slow rhythms are the norm.

\begin{figure}[H]
 \centerline{\framebox{
 \includegraphics[width=0.95\columnwidth]{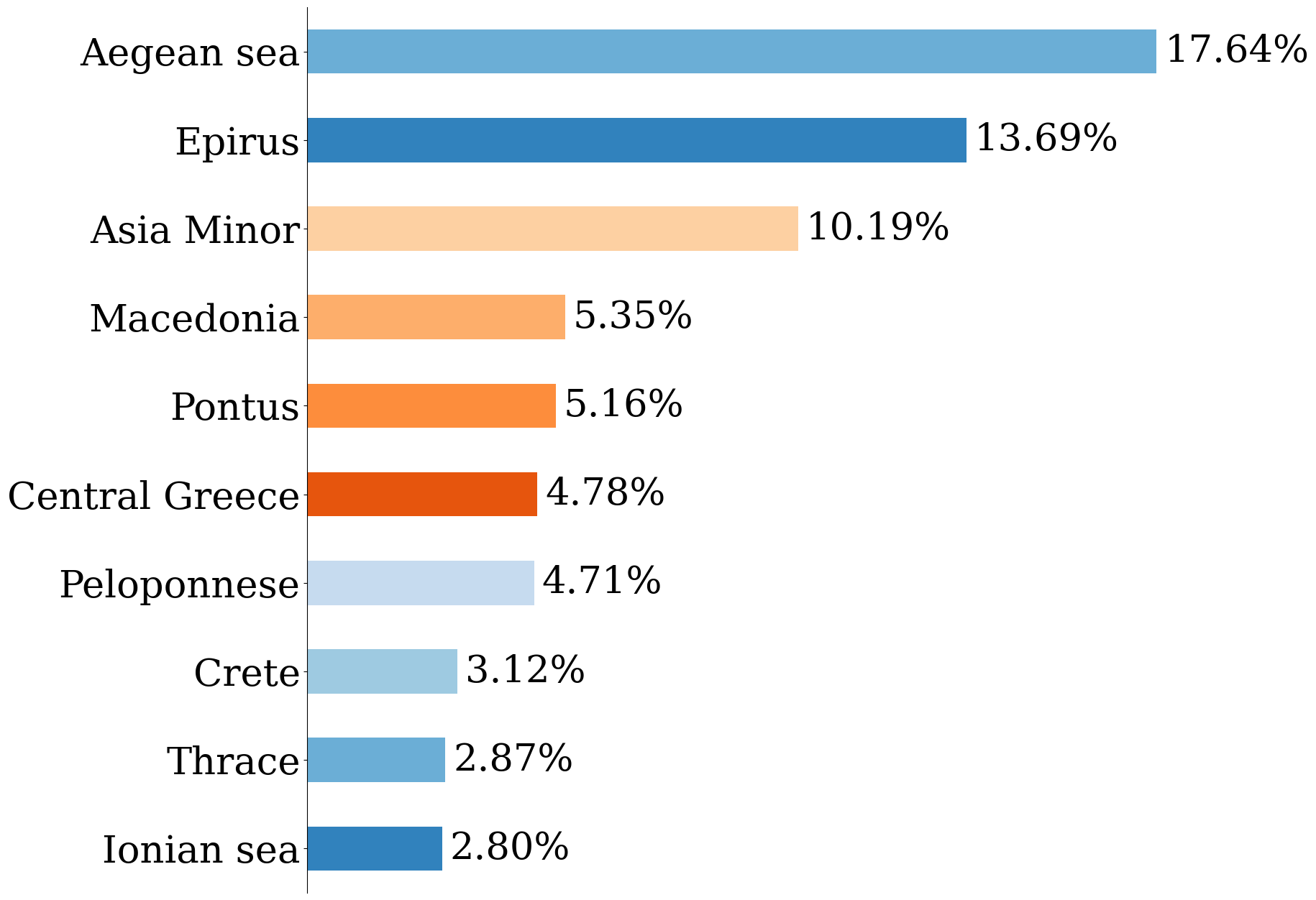}}}
 \caption{Relative frequencies of the most represented regions in the dataset.}
 \label{fig:places}
\end{figure}

The ``place'' (of origin) metadata category can be annotated with (i) a single label when the region from which a song derives is known, (ii) two labels when both region and a specific place are known and (iii) ``None'' denoting that there is not a specific place of origin for this piece. As an example, a music piece can be annotated with the region ``Aegean sea'' or with both ``Aegean sea'' and ``Naxos'', an island of the Aegean sea, if this knowledge is available. Specifically, from the 81 unique places in the dataset, 20 are regions and only half of them include the remaining 61. The most represented regions can be seen in \figref{fig:places}.

\begin{figure}[H]
 \centerline{\framebox{
 \includegraphics[width=0.95\columnwidth]{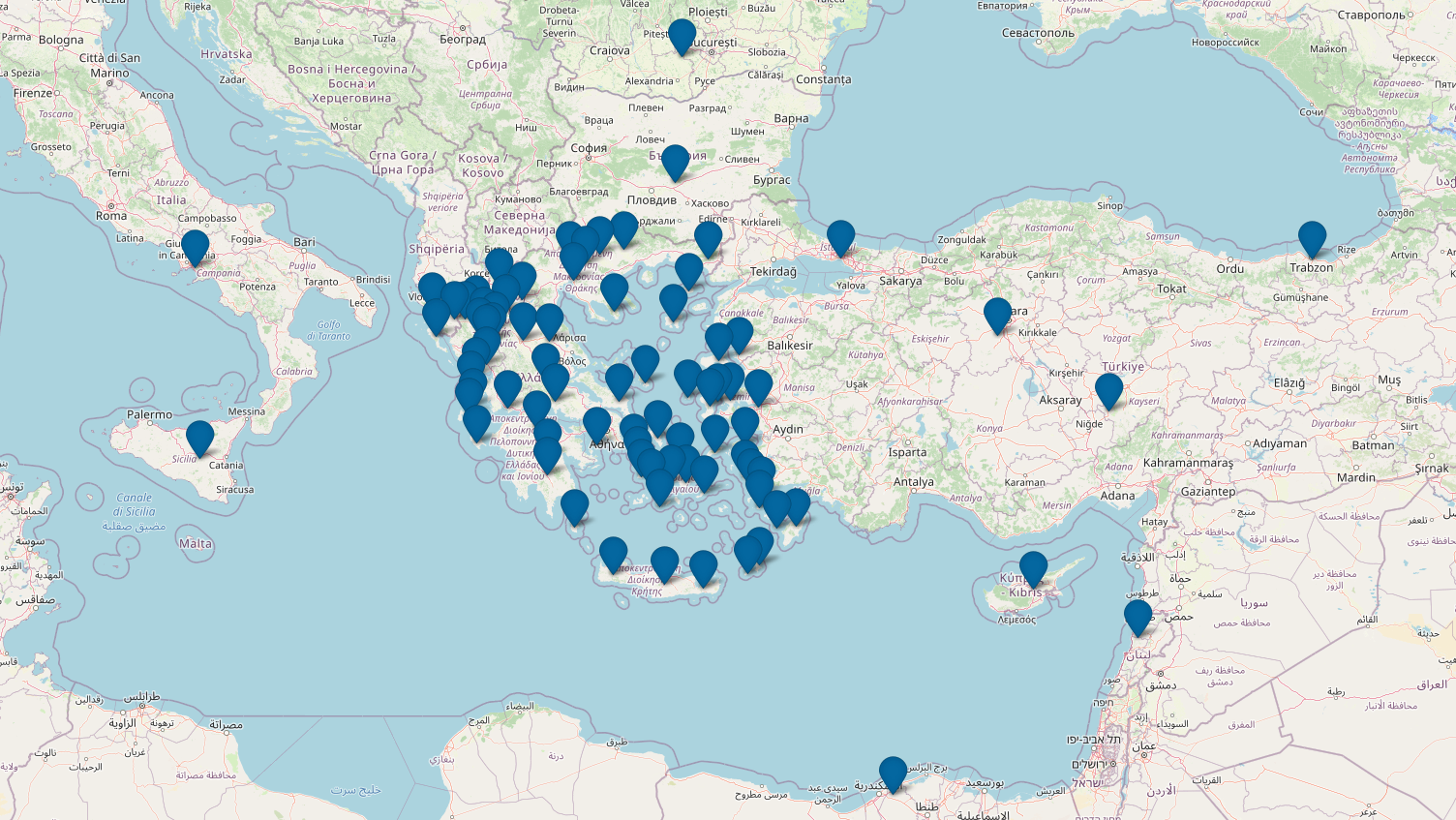}}}
 \caption{Map of all the places of the dataset.}
 \label{fig:map}
\end{figure}

The exact latitude and longitude of each place is also available at the ``coordinates'' column. The  music pieces that do not have an explicit place of origin, such as the ones that belong to the ``laiko'' genre, are accounted for approximately 23\% of the total. 
\figref{fig:map} shows the location of the 81 places that exist in the dataset. We may notice the constant ability of music to excess the borders; places where Greek culture thrived in the past and neighboring countries that share the same tradition, form a mosaic of people that communicated freely with each other in a musical way that has reached towards us.

Analogous connections, like the ones between genres and places, one expects to be observed between places and instruments as well. Indeed, for over 100 years, the established music ensembles of Greek traditional music are generally two, namely (i) those with the violin as leading instrument (often substituted by lyra and santouri), which have a greater presence in island Greece and (ii) those with the klarino (Greek clarinet) as the leading instrument, which is dominant in the mainland. 

\begin{figure}[H]
 \centerline{\framebox{
 \includegraphics[width=0.95\columnwidth]{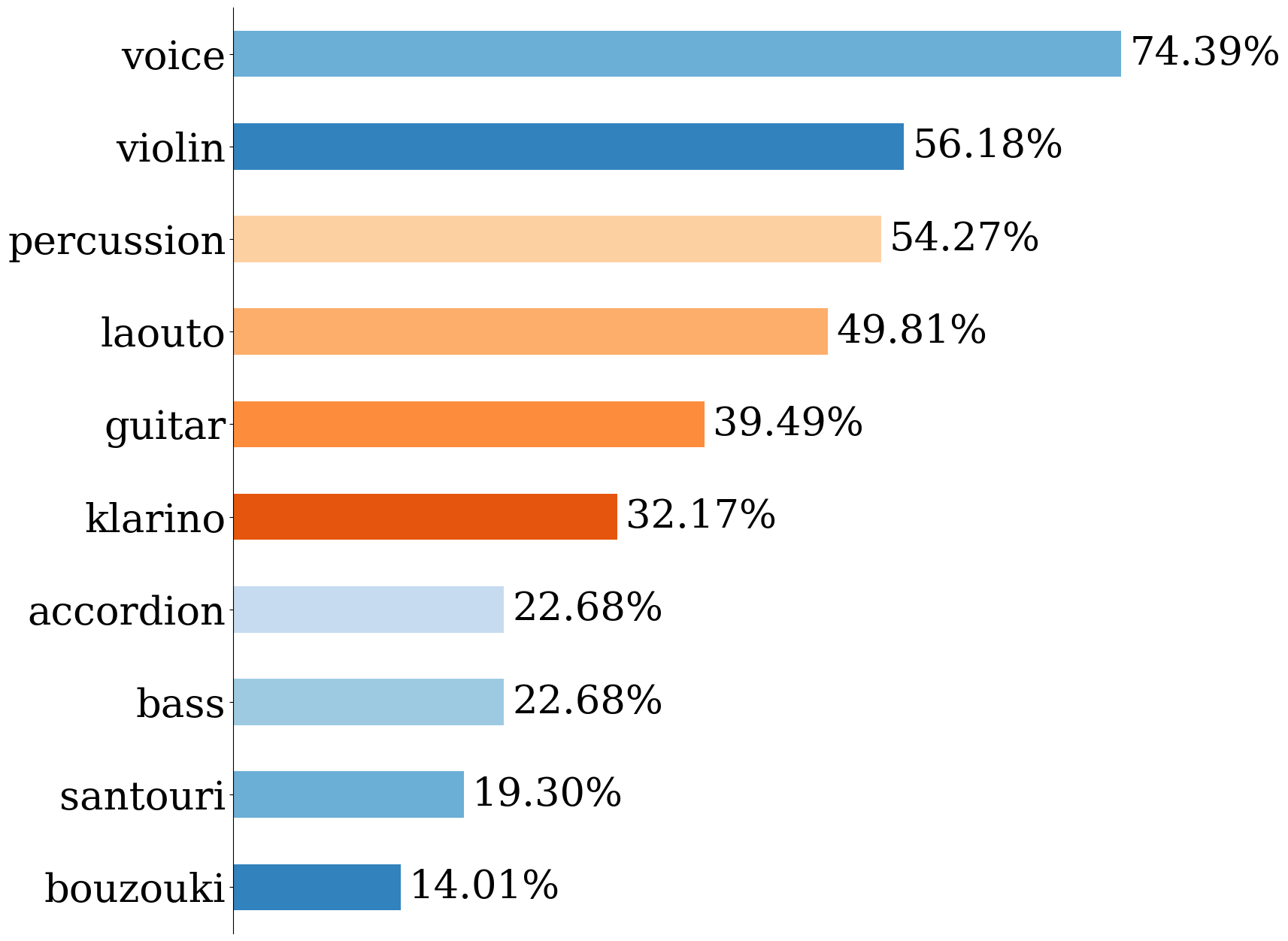}}}
 \caption{Relative frequencies of the most common musical instruments in the dataset.}
 \label{fig:instruments}
\end{figure}

In the popular and modern music domain, there is a great variety of instruments, but in most cases bouzouki, guitar, accordion and bass are common members of a laiko or rebetiko music ensemble. In the traditional music groups, the percussion and the laouto (Greek lute) are permanent companions of the leading instruments, offering melodic-harmonic background and rhythmic support. Of course, voice holds the main role in all kinds of performances.

In \figref{fig:instruments} one can see the frequencies of the most popular instruments in the dataset. Singing voice is evident in almost 75\% of it and instruments like violin, percussion and laouto, that have presence in both islands and mainland as well, are following.

\begin{figure}[H]
 \centerline{\framebox{
 \includegraphics[height=10.4cm, keepaspectratio]{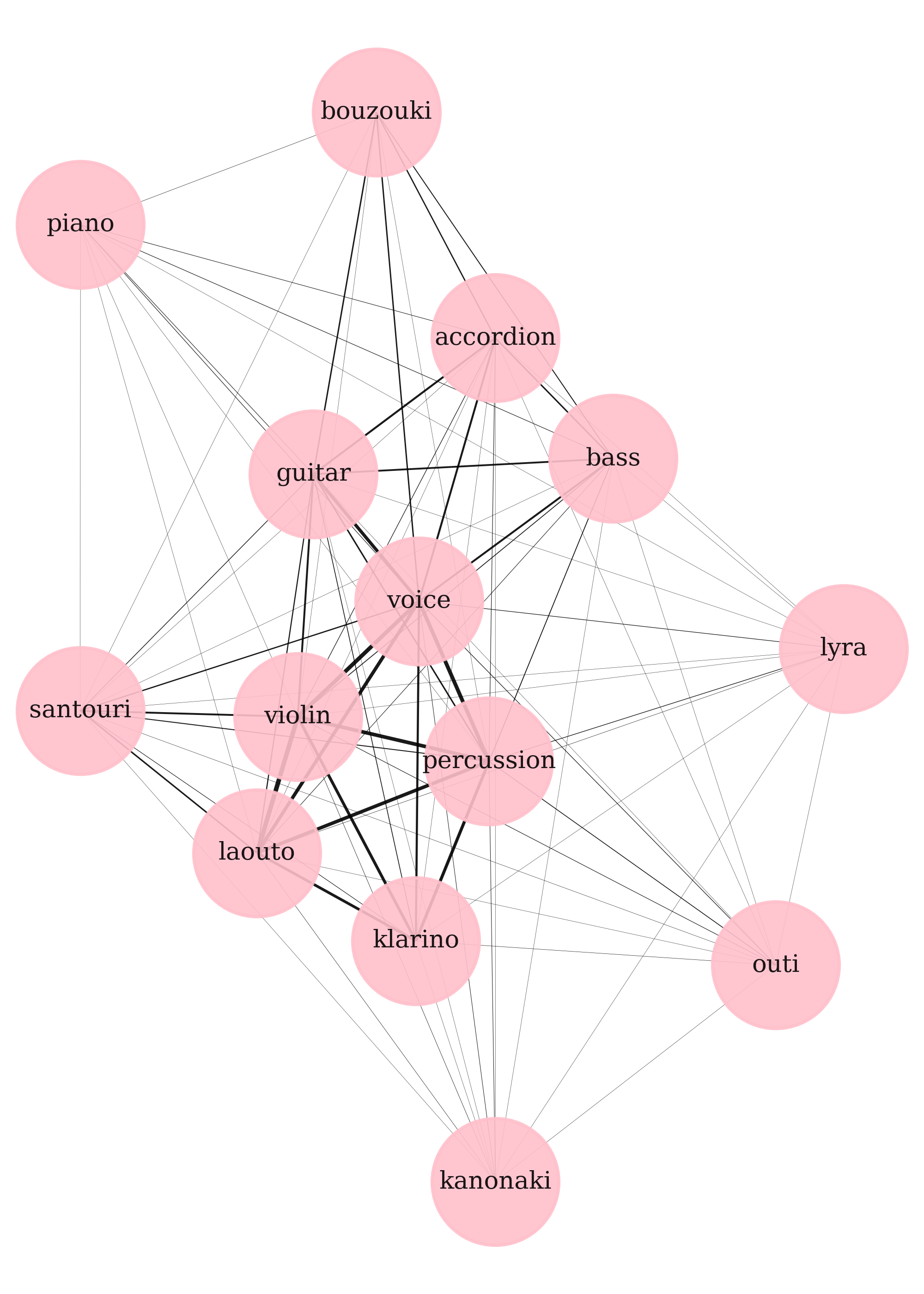}}}
 \caption{Graph that shows the co-occurrences of the fourteen most common musical instruments. The edge width is proportional to the number of music pieces a pair of instruments co-occur.}
 \label{fig:graph}
\end{figure}

With regards to the music ensembles, it should be noted that 296 unique groups of instruments exist in the dataset, with the one constituted by voice, violin, percussion, laouto and klarino being by far the most popular by participating in the performance of around 12\% of the music pieces. The co-occurrences of the most popular instruments can be seen in \figref{fig:graph} where the width of the graph edges is proportional to the number of pieces a pair of instruments co-occur in the dataset.

Sample rows of the dataset can be seen in \tabref{tab:rows}. The dataset along with the baseline classification methods and the trained models are available online.\footnote{\texttt{https://github.com/pxaris/lyra-dataset}}

\begin{table*}[ht!]
 \begin{center}
 \begin{scriptsize}
 \texttt{
 \begin{tabular}{|c|c|c|c|c|c|c|c|c|}
  \hline
  \textbf{id} & \textbf{instruments} & \textbf{genres} & \textbf{place} & \textbf{coordinates} & \textbf{is-danced} & \textbf{youtube-id} & \textbf{start-ts} & \textbf{end-ts} \\
  \hline
   alexandra & \begin{tabular}{@{}c@{}}voice|violin|\\ percussion|\\ laouto|\\ klarino \end{tabular} & \begin{tabular}{@{}c@{}}Traditional|\\ Epirotic \end{tabular} & \begin{tabular}{@{}c@{}}Epirus|\\ Zagori \end{tabular} & \begin{tabular}{@{}c@{}}39.8648|\\ 20.9284 \end{tabular} & 0 & qrOwc1mLFUk & 749 & 927\\
  \hline
  choros-tik & \begin{tabular}{@{}c@{}}percussion|\\ lyra \end{tabular} & \begin{tabular}{@{}c@{}}Traditional|\\ Pontian \end{tabular} & Pontus & \begin{tabular}{@{}c@{}}40.9883|\\ 39.7270 \end{tabular}  & 1 & Aws0Y3aLaIs & 1731 & 1886\\
  \hline
  \begin{tabular}{@{}c@{}}agiothodo- \\ ritissa \end{tabular} & \begin{tabular}{@{}c@{}}voice|violin|\\ santouri|\\ percussion|\\ laouto|guitar \end{tabular} & Rebetiko & None & None & 1 & 0cj8BNcAhg4 & 2632 & 2853\\
  \hline
  \begin{tabular}{@{}c@{}}einai-arga- \\ poly-arga \end{tabular} & \begin{tabular}{@{}c@{}}voice|piano|\\ guitar|bass|\\ bouzouki|\\ accordion \end{tabular} & Laiko & None & None & 0 & zkoqg3VRVLA & 2365 & 2614\\
  \hline
  \end{tabular}
 }
\end{scriptsize}
\end{center}
 \caption{Sample rows of the dataset. Pipe ``|'' is used to separate values in a field.}
 \label{tab:rows}
\end{table*}

The shared metadata should be considered as the version 1.0 of the dataset. In the next versions, it will be evolved towards two main directions, namely (i) the incorporation of more metadata categories such as annotations according to the content of the lyrics, the lyrics themselves as well as information about the types of the dances that occur and (ii) the addition of more music pieces by following the same process either for next episodes of the same documentary series or for other series that have a similar theme.

%% file: 3_application.tex
\section{Baseline Classification}\label{sec:application}

The audio recording of each music piece is represented using a Mel-scaled Spectrogram (mel-spectrogram): this is  a spectrogram whose frequencies are converted to the mel scale according to the equation:
\begin{align}
m = 2595\log_{10}(1+\frac{f}{700}) = 1127\ln(1+\frac{f}{700})\label{eq3}
\end{align} 
where \textbf{m} is the frequency in Mels and \textbf{f} is the frequency in Hz. Mel-spectrograms are calculated per fix-sized segment duration of 10 seconds. A non-overlapping window of 50 milliseconds has been applied, therefore the Mel-spectrogram size is $200$ windows $\times$ $128$ frequency bins. 

As a baseline classification approach, each 10-second Mel-spectrogram is classified to the aforementioned tasks (genre, place and instruments) using a Convolutional Neural Network (CNN). CNNs have been widely used in general audio \cite{papakostas2018speech}, speech \cite{huang2014speech, choi2017convolutional} and music \cite{ashraf2020globally} classification tasks. In particular, we have adopted the following architecture: 4 convolutional layers of 5 $\times$ 5 kernels, single stride, and max pooling of size 2. The number of convolutional kernels (channels) are for the first layer  32, for the second 64, for the third 128 and for the fourth 256. The final output of the convolutional layers is passed through 3 fully connected layers, with the first having an output dimension of 1024, the second 256 and the third equal to the number of classes. 

Note that fix-sized duration of segments is necessary, since audio recordings do not share the same size. Adopting a much longer segment would require zero padding for several mel-spectrograms and probably more CNN parameters. In addition, splitting the song into non-overlapping segments achieves some type of data augmentation. For two of the adopted classification tasks (namely genre and place), we have trained the CNNs using a multiclass, single-label setup, while for the instrument task, which is multi-label, we have trained multiple binary CNNs, one for each instrument, which have been evaluated separately. 

After the training and validation procedure of each of the aforementioned CNNs, final testing was applied on the respective test recordings. For the test set, to avoid spreading pieces from the same broadcast across data splits, we separate training and test data on an episode level. From the 74 unique episodes, we randomly split 20\% of them and use all the music pieces they include, namely 330, to form the test data. Obviously, this final testing needs to be carried out on a ``song-level'', not a 10-sec segment level. Towards this end, a simple aggregation method was adopted, by just averaging the posteriors of the individual segment decisions of the CNNs. This aggregated estimate was used as the final prediction and evaluated in the final testing.

%% file: 4_results.tex
\section{Results}\label{sec:results}

The performance results during the training of the baseline classifiers are shown in Table \ref{tab:train}, computed on a validation subset that corresponds to 20\% of the segments. For the multi-label task (instrument recognition), we show the F1 metric for each binary subtask separately, while for the single-label tasks (genre and place) we show the overall macro F1 for all classes. We remind that this evaluation is performed on the validation split of the 10-second data.

\begin{table}[H]
 \begin{center}
 \begin{tabular}{|c|c|c|}
  \hline
  \textbf{Classifiers type} & \textbf{Task} & \textbf{F1} (\%) \\
  \hline
  \multirow{2}{*}{\begin{tabular}{@{}c@{}} Multiclass \\
  classifiers \end{tabular}} & Genre & 82.3 \\ 
                             & Place & 85.5 \\
  \hline
  \multirow{6}{*}{\begin{tabular}{@{}c@{}} Binary \\
  classifiers \\ for \\ Instruments \end{tabular}} & voice & 75.8 \\
                                                   & violin & 86.3 \\
                                                   & percussion & 97.1 \\
                                                   & laouto & 96.7 \\
                                                   & guitar & 80.2 \\
                                                   & klarino & 95.0 \\
  \hline
 \end{tabular}
\end{center}
 \caption{F1 scores of the classifiers on the 10-second segments of the training data using a 20\% validation subset.}
 \label{tab:train}
\end{table}

As soon as the 10-second classifiers are trained, they are applied on the whole recordings of the testing data, and a simple majority aggregation is performed to extract the final decision, as described in the previous Section. For the instrument classification task, we compute the Area Under the Curve (AUC) metric per label (binary classification subtask). The results are shown in Table \ref{tab:test_1}.

\begin{table}
 \begin{center}
 \begin{tabular}{|c|c|}
  \hline
  \textbf{Instrument} & \textbf{AUC} (\%) \\
  \hline
  voice & 68.9 \\
  \hline
  violin & 85.2 \\
  \hline
  percussion & 95.1 \\
  \hline
  laouto & 93.8 \\
  \hline
  guitar & 73.5 \\
  \hline
  klarino & 90.9 \\
  \hline
 \end{tabular}
\end{center}
 \caption{Area Under the Curve (AUC) scores of the instrument classifiers on the test data.}
 \label{tab:test_1}
\end{table}

\begin{figure}
 \centerline{\framebox{
 \includegraphics[width=0.9\columnwidth]{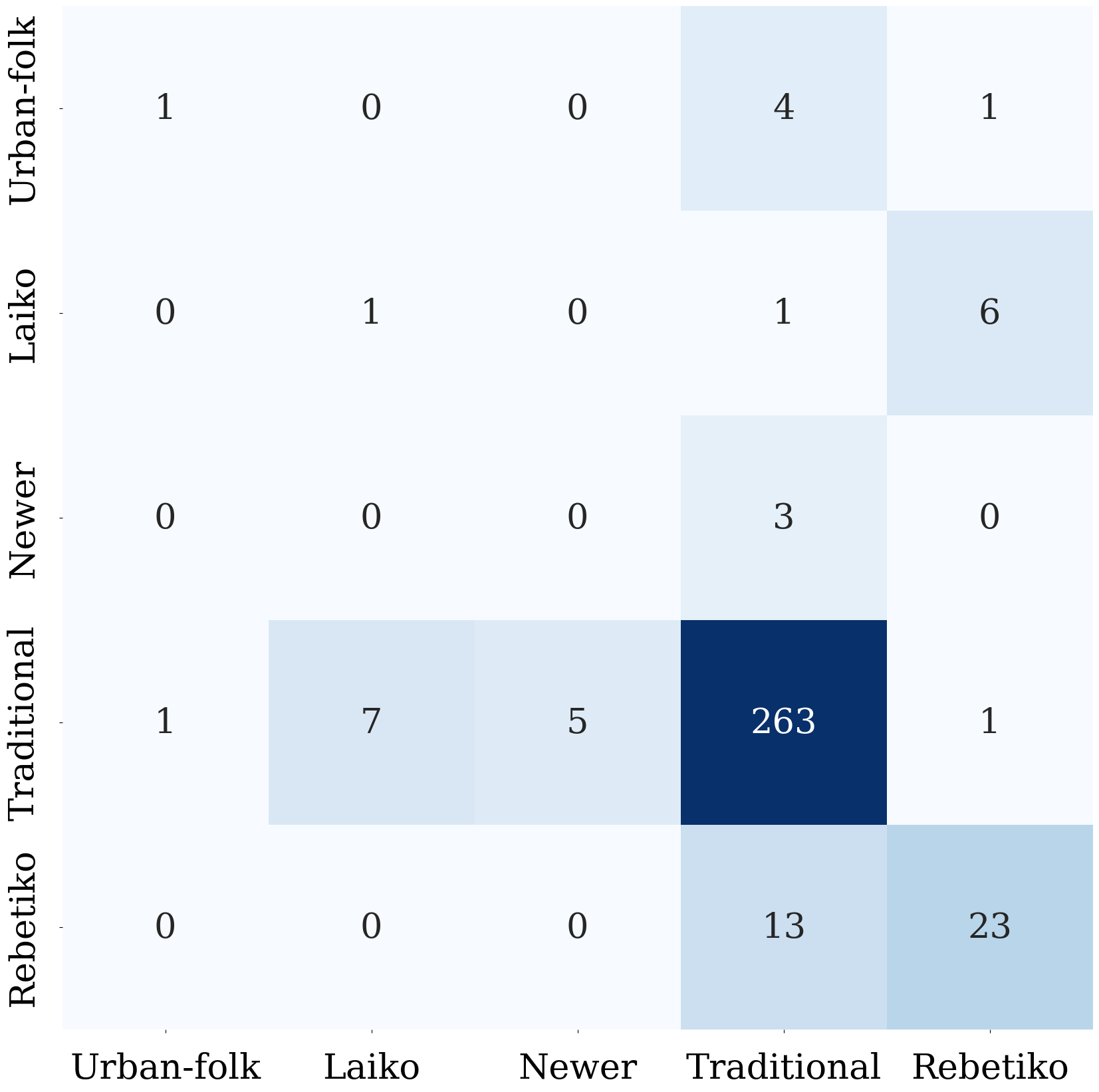}}}
 \caption{Confusion matrix for ``genre'' classes on the test data. \textbf{Macro F1-}score: \textbf{39.9}\% and \textbf{Micro F1-}score: \textbf{87.2}\%.}
 \label{fig:cm_genres}
\end{figure}

\begin{figure}
 \centerline{\framebox{
 \includegraphics[width=0.9\columnwidth]{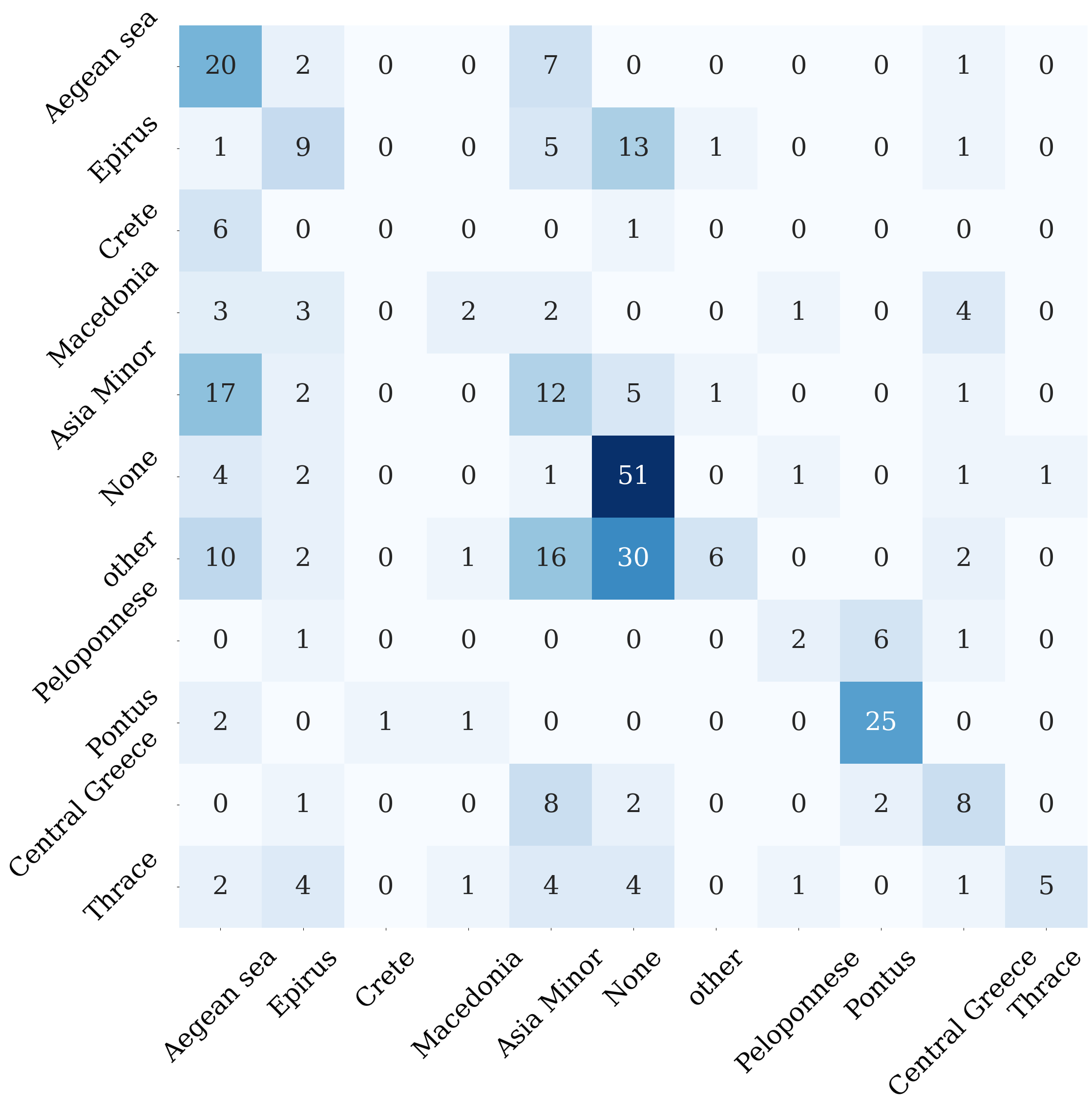}}}
 \caption{Confusion matrix for ``place'' classes on the test data. \textbf{Macro F1-}score: \textbf{34.4}\% and \textbf{Micro F1-}score: \textbf{42.4}\%.}
 \label{fig:cm_places}
\end{figure}

Finally, the confusion matrices along with the respective F1 measures for the multiclass, single-label classification tasks of ``genres'' and ``places'' are shown in Figures \ref{fig:cm_genres} and \ref{fig:cm_places}. All genres have been taken into account for the respective classification, but not the sub-genres. On ``places'' task, the pieces have been classified to the 10 most common regions (including ``None'') plus the ``other'' category for the remaining. Genres classifier macro F1-score is 39.9\% and micro F1-score is 87.2\%, while for the places classifier the macro F1-score is 34.4\% and the micro F1-score is 42.4\%.  

%% file: 5_discussion.tex
\section{Discussion}\label{sec:discussion}

A reason that the ``voice'' classifier has lower performance compared to the other ones may be of a musicological character. Indeed, while the presence of the rest of the instruments can depend significantly on the music style of a piece, the same does not apply to ``voice'' as it is the dominant musical instrument in any genre. Given the fact that the binary classifier is trained to recognize an instrument (evident in a part of a music piece) in each of the 10-second segments, regardless if it is present on it or not, we expect to move towards a space with latent musical features such as the music style, where ``voice'' may not be as discriminative as the rest of the instruments are.

With regards to confusion matrices, the misclassifications can be either due to imbalance between classes or statistical correlations across them. Specifically, for the ``places'' task, the confusions between regions that are geographically near may be justifiable, while for ``genres'' task the imbalance between the classes seems to have significantly affected the performance of the model at the least represented ones.

We intend to improve models performance and conduct a more in-detail analysis in order to examine hypotheses such as the aforementioned ones in a follow-up study.

%% file: 6_conclusions.tex
\section{Conclusions}\label{sec:conclusions}

Greek traditional and folk music integrates components of Eastern and Western idioms, providing interesting research directions in the field of computational ethnomusicology. This paper presents ``Lyra'', a dataset of 1570 traditional and folk Greek music pieces that includes audio and video (timestamps and links to YouTube videos), along with annotations that describe aspects of particular interest for this dataset, including instrumentation, geographic information and labels of genre and subgenre, among others. The advantage of this dataset is that the entire content is harvested from web resources of a Greek documentary series that was produced by academics with specialization in this music and, therefore, includes high-quality and rich annotations extracted from the content of the shows. Additionally, the production of recordings and video material is professional-level, providing a common ground in terms of audio quality. Three baseline audio-based classification tasks are performed, namely instrument identification, place of origin and genre classification.


The presented results indicate that specialized tasks, that use the audio signal, can potentially provide valuable insight about several aspects of this music. The combination of video and audio signals allows possible experimentation on methods that process multimodal data. The Lyra dataset includes material that readily allows MIR tools to be employed for reaching valuable musicological results. This dataset can also, potentially, foster the expansion of the MIR methods altogether.

%% file: 7_acknowledgements.tex
\section{Acknowledgements}\label{sec:acknowledgements}

The authors would like to thank Professor Lambros Liavas, the producer, researcher and presenter of the documentary series "To Alati tis Gis - Salt of the Earth". His work not only provides a fruitful source of knowledge and art but also acts as an inspiration for new researchers. This work would not have been possible without the help of the volunteer annotators. A big thank to the students of the Department of Music Studies of National and Kapodistrian University of Athens. Finally, we would like to thank the reviewers for their valuable and constructive comments.